\def\be{\begin{equation}}
\def\ee{\end{equation}}
\def\bea{\begin{eqnarray}}
\def\eea{\end{eqnarray}}

\documentclass[aps,prl,twocolumn,amsmath,amssymb]{revtex4}
\usepackage{epsfig,graphicx}
\usepackage{bm}

\begin{document}


\title{Persistent Rabi oscillations probed via low-frequency noise correlation}

\author{Alexander N. Korotkov}
\affiliation{Department of Electrical Engineering, University of
California, Riverside, CA 92521, USA
}

\date{\today}


\begin{abstract}
The qubit Rabi oscillations are known to be non-decaying (though
with a fluctuating phase) if the qubit is continuously monitored in
the weak-coupling regime. In this paper we propose an experiment to
demonstrate these persistent Rabi oscillations via low-frequency
noise correlation. The idea is to measure a qubit by two detectors,
biased stroboscopically at the Rabi frequency. The low-frequency
noise depends on the relative phase between the two combs of biasing
pulses, with a strong increase of telegraph noise in both detectors
for the in-phase or anti-phase combs. This happens because of
self-synchronization between the persistent Rabi oscillations and
measurement pulses. Almost perfect correlation of the noise in the
two detectors for the in-phase regime and almost perfect
anticorrelation for the anti-phase regime indicates a presence of
synchronized persistent Rabi oscillations. The experiment can be
realized with semiconductor or superconductor qubits.
\end{abstract}


\maketitle

The puzzle of the quantum state collapse due to measurement
\cite{Wheeler-Zurek} is becoming accessible for the experimental
study in solid state systems. Three experiments on non-projective
collapse \cite{Katz,Bertet} have been recently realized with
superconducting qubits. These experiments (as well as the experiment
proposed in the present paper) touch upon the most intriguing
property of quantum measurement: the presence of a ``spooky''
quantum back-action \cite{EPR}, which changes the system to agree
with the observation, and cannot be explained in a realistic way,
i.e.\ by using the Schr\"odinger equation.

    The quantum coherent (Rabi) oscillations in solid-state qubits
are usually measured in an ensemble-averaged way
\cite{Rabi-exp,Rabi-semicond} and decay within a short timescale,
even though it can be much longer than the oscillation period.
However, for a continuous weak measurement of a single qubit, the
Rabi oscillations are non-decaying and can in principle be monitored
in real time, as follows, e.g., from the quantum Bayesian formalism
\cite{Kor-99}, which is generally similar to the formalism of
quantum trajectories \cite{quant-traj}. Persistence of the Rabi
oscillations in this case is due to the quantum back-action, which
tends to increase the amplitude of the oscillations to 100\%, thus
competing against decoherence. The persistent Rabi oscillations lead
to the spectral peak of the detector signal at the Rabi frequency
\cite{Rabi-peak,Kor-Rabi-peak}, which has been recently observed
experimentally \cite{Bertet} (see also \cite{peak-exp-old}). In the
present paper we will discuss another way of demonstrating these
oscillations.

    For definiteness let us discuss a ``charge'' qubit made of a double
quantum dot (DQD) populated by a single electron, the location of
which is continuously measured by a nearby quantum point contact
(QPC). Analogous setups can be realized with spin-based or
superconducting qubits. The continuous qubit evolution due to the
quantum ``informational'' back-action can in principle be verified
in a direct experiment \cite{Kor-99}; however, it would require
high-bandwidth recording of the detector signal (including shot
noise) and fast qubit manipulation, that is still a big challenge
for a real experiment. A simpler way to study the back-action is to
measure the qubit by two detectors \cite{Kor-2q-exp}, so that the
first (short) measurement causes a partial collapse of the qubit
state, and then after a controllable qubit evolution the second
detector measures the resulting state. Performing the experiment
many times and selecting a certain result of the first measurement,
it is possible to find the back-action evolution experimentally and
compare it with the theory. The same idea with a different
post-processing (selecting the result of the second measurement) has
been recently used to propose an experiment on weak values
\cite{Blanter}.

    The proposal of Ref.\ \cite{Kor-2q-exp} still suffers from
very weak signals produced by two single-shot measurements. An
obvious way to increase the signal is to average it over a long comb
of measurement pulses, but in this case the selection of a certain
result becomes impossible. Fortunately, there is a way to overcome
this dilemma by combining the ideas of two-detector measurement
\cite{Kor-2q-exp}, persistent Rabi oscillations
\cite{Kor-99,Rabi-peak,Kor-Rabi-peak}, and stroboscopic quantum
non-demolition (QND) measurement \cite{Braginsky-book,kicked}.

\begin{figure}[tb]
  \centering
\includegraphics[width=7.0cm]{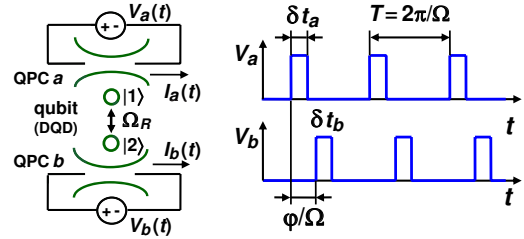}
  \caption{Analyzed system: a double-quantum-dot qubit measured
by two QPC detectors, which are biased by combs of short voltage
pulses with frequency $\Omega$ coinciding with the Rabi frequency
$\Omega_R$. }
  \label{fig1}
\end{figure}

   We propose to use the following setup (Fig.\ 1). A qubit is
measured by two detectors, which are biased with two combs of short
pulses, so that between the pulses the qubit undergoes free
evolution due to the Rabi oscillations. The frequency of pulses
$\Omega$ coincides with the Rabi frequency $\Omega_R$ (one pulse per
period in each detector) to realize the QND regime \cite{kicked}.
When the two combs are not shifted in time relative to each other,
the phase of the Rabi oscillations is attracted to one of the two
stable values, corresponding to the qubit being in either localized
state $|1\rangle$ or $|2\rangle$ at the time of measurement. This
happens because of the usual collapse in the QND frame
\cite{Braginsky-book,kicked} and is somewhat similar to what happens
with a parametrically excited swing. However, because of various
imperfections (extra decoherence, etc.) there will be switching
between the two stable regimes, which leads to the telegraph noise
in the currents through both detectors. Even if the experimental
measurement bandwidth is not wide enough to resolve the switching
events (which is likely for a present-day experiment), the telegraph
noise is measurable at low frequency via its spectral density, which
greatly exceeds the shot noise.

    The telegraph noise originates from the presence of two
quasi-stable regimes of oscillations because of the QND measurement.
However, for a significant phase difference $\varphi$ between the
two combs of the measurement pulses, the measurement is no longer
QND, and the telegraph noise disappears, so that the low frequency
noise reduces to a much smaller shot noise. For $\varphi\approx\pi$
the QND regime is restored again and the telegraph noise reappears.
So the telegraph noise is maximum at both $\varphi=0$ and
$\varphi=\pi$; however, for $\varphi=0$ it is almost fully
correlated \cite{antisym} in the two detectors (because in the
stable regime both detectors see the same qubit state: either
$|1\rangle$ or $|2\rangle$), while for $\varphi=\pi$ the detector
noises are almost fully anticorrelated (because when one detector
measures state $|1\rangle$, the other detector measures $|2\rangle$
half a period later). Experimental observation of such noise
correlation/anticorrelation would show the presence of persistent
Rabi oscillations.

    For the quantitative analysis let us use the quantum Bayesian
formalism \cite{Kor-99} and represent the QPC currents $I_a(t)$ and
$I_b (t)$ as ($n=a$ or $n=b$)
    \be
    I_{n}(t) = [I_{0,n} + (\Delta I_{n}/2) \, z(t)] \,
    f_{n}(t)+ \xi_{n}(t) \, \sqrt{|f_{n}(t)|} ,
    \label{I-n}\ee
where $z=\mbox{Tr}(\sigma_z\rho )$ is the measured $z$-component of
the qubit Bloch vector, $f_n(t)$ is the dimensionless shape of the
comb of measurement pulses for $n$th detector ($f$ is proportional
to the QPC voltage), $I_{0,n}$ and $\Delta I_n$ are the detector
average current and response for $f_n=1$, and $\xi_n(t)$ is the shot
noise with the (one-sided) spectral density $S_n$. We assume zero
temperature. For rectangular measurement pulses we use $f_a(t)=1$ if
$|t-lT|<\delta t_a/2$ and $f_a(t)=0$ otherwise, where $\delta t_a$
is the pulse duration, $T=2\pi/\Omega$ is the comb period, and $l$
is an integer. Similarly, $f_b(t)=1$ if
$|t-lT-\varphi/\Omega|<\delta t_b/2$; this describes the comb of
pulses of duration $\delta t_b$ with the same frequency $\Omega$ but
shifted by the phase $\varphi$. The qubit Hamiltonian $H_{\rm
qb}=(\Omega_R/2)\,\sigma_x$ describes Rabi oscillations with the
frequency $\Omega_R$ about the $x$-axis, and we also assume pure
dephasing of the qubit (not related to the measurement) with rate
$\gamma$. In order to use the Markovian approximation, we assume
sufficiently high QPC voltages during the pulses \cite{Kor-99}, and
also assume $|\Delta I_n|\ll |I_{0,n}|$.

    The spectral densities of two detector noises $S_{aa}(\omega )$ and
$S_{bb}(\omega )$ as well as the cross-correlation noise
$S_{ab}(\omega )$ can be found via the Fourier transform
$S_{nm}(\omega)=2\int_{-\infty}^{\infty} \overline{K_{nm} (\tau,t_0
)}\,e^{-i\omega\tau}d\tau$, where $K_{nm}(\tau,t_0)=\langle
I_m(t_0+\tau)I_n(t_0)\rangle - \langle I_m(t_0+\tau)\rangle \,
\langle I_n(t_0)\rangle $ is the correlation function and the
averaging $\overline{K_{nm} (\tau,t_0 )}$ is over $t_0$ within one
period $T$; the averaging is necessary because of periodic time
dependence $f_n(t)$ in Eq.\ (\ref{I-n}). To find $K_{nm} (\tau,t_0
)$ for $\tau
>0$ we use the method developed in \cite{Kor-Rabi-peak}, which
essentially follows from the quantum regression theorem
\cite{Milburn-Walls}. Expressing this correlator as $K_{nm}
(\tau,t_0 )=(\Delta I_n\Delta I_m/4)f_m(t_0+\tau)f_n(t_0)
K^{zz}(\tau,t_0)$, we calculate the operator-symmetrized
$zz$-correlator $K^{zz}(\tau,t_0)$ as
    \be
K^{zz}(\tau,t_0)= \sum_{i=1,2} \langle i| \rho(t_0) |i\rangle
 \langle i| \sigma_z |i\rangle \,
 \mbox{Tr}[\sigma_z\rho^{|i\rangle,t_0}(t_0+\tau)],
    \label{K-zz}\ee
where $\rho(t_0)$ is the qubit density matrix at time $t_0$
[$\langle i| \rho(t_0) |i\rangle=1/2$ because of the symmetry],
while $\rho^{|i\rangle,t_0}(t_0+\tau)$ is the density matrix at time
$t_0+\tau$ for the qubit starting at time $t_0$ in the state
$|i\rangle$, which is an eigenstate of $\sigma_z$. Here we should
assume the ensemble-averaged qubit evolution, for which the
measurement process is represented by pure dephasing with rates
$f_n(t)\gamma_n^{\rm meas}$, where $\gamma_n^{\rm meas}= (\Delta
I_n)^2/4S_n$. Notice that the same technique works also for two
detectors measuring different observables: one of them defines the
starting eigenstates, and the other one enters into the trace. While
so far $\tau>0$ was assumed, we find the correlation function at
negative $\tau$ using the symmetry $K_{nm}(\tau,t_0)=K_{mn}(-\tau,
t_0+\tau)$, and also add the shot noise contribution
$\delta_{nm}\delta(\tau )f_n(t_0)S_n/2$ near $\tau=0$. Besides this
quantum method, we also use below the language of a simple
semiclassical analysis.

    Let us first assume short measurement pulses, $\delta
t_{a,b}\ll T=2\pi/\Omega$, exactly matched frequency,
$\Omega=\Omega_R$, almost no phase shift, $|\varphi| \ll 1$, and
almost negligible extra dephasing, $\gamma \lll \Omega_R$. Then we
have usual stroboscopic QND measurement \cite{Braginsky-book,kicked}
insensitive to the free evolution, and therefore the qubit state
eventually collapses to either $|1\rangle$ or $|2\rangle$ at the
measurement moments. This obviously leads to non-decaying Rabi
oscillations with 100\% amplitude, which are phase-locked with the
measurement combs (though with a random choice of the stable phase).
The synchronization  happens within the QND collapse timescale
$t\sim t_{\rm col} = T/(M_a+M_b)$, where $M_n =\gamma_n^{\rm
meas}\delta t_n=\delta t_n(\Delta I_n)^2/4S_n$. We assume $M_n \ll
1$, so that $t_{\rm col} \gg T$, while $\gamma_{a,b}^{\rm meas}T$
are not necessarily small. In the ideal QND case the phase of the
Rabi oscillations is fixed forever after this gradual collapse;
however, in a realistic case there will be switching between the two
regimes (state $|1\rangle$ or $|2\rangle$ at the measurement
moments) with a calculated below rate $\Gamma_S$, the same for both
switching processes because of the symmetry. If we assume rare
switching, $\Gamma_S \, t_{\rm col} \ll 1$, then the detector
current $I_n$ averaged over a coarse graining timescale longer than
$T$ and $t_{\rm col}$, switches between the two levels, $(I_{0,n}
\pm \Delta I_n/2)(\delta t_n/T)$, thus producing the telegraph
noise. Therefore, the noise spectral density $S_{nn}(\omega)$ at
frequencies $\omega \ll t_{\rm col}^{-1} \ll \Omega$, is
    \be
    S_{nn}(\omega )= \left(\frac{\delta t_n}{T} \right) ^2
    \frac{(\Delta I_n)^2/2\Gamma_S}{1+(\omega/2\Gamma_S)^2} +
    \frac{\delta t_n}{T} S_n ,
    \label{S-nn}\ee
where the term $(\delta t_n/T)S_n$ is due to the shot noise. It is
easy to see that at low frequency, $\omega \ll \Gamma_S$, the ratio
of the telegraph and shot noise contributions $(\delta t_n/T)(\Delta
I_n)^2/2\Gamma_S S_n \simeq 1/t_{\rm col}\Gamma_S$ is always large
in our case.

    For the phase shift $\varphi \approx \pi$ the QND regime is still
realized, and therefore Eq.\ (\ref{S-nn}) for each detector noise is
still valid. However, since the detectors now measure the opposite
qubit states, the cross-correlation noise changes sign,
   \be
    S_{ab}(\omega )= \pm \frac{\delta t_a \, \delta t_b}{T^2}
    \frac{\Delta I_a \Delta I_b/2\Gamma_S}{1+(\omega/2\Gamma_S)^2} ,
    \label{S-ab}\ee
where ``$+$'' sign is for $\varphi \approx 0$ and ``$-$'' is for
$\varphi \approx \pi$. The noise correlation factor
$S_{ab}(0)/\sqrt{S_{aa}(0)S_{bb}(0)}$ is obviously close to $\pm 1$,
describing almost full correlation/anticorrelation, when the shot
noise term in Eq.\ (\ref{S-nn}) is much smaller than the telegraph
noise.

    To find the switching rate $\Gamma_S$, we calculate the
``propagators''    $\rho^{|i\rangle,t_0}(t_0+\tau)$  in Eq.\
(\ref{K-zz}), essentially rederiving Eqs.\ (\ref{S-nn}) and
(\ref{S-ab}) in the fully quantum way. Because of assumed weak
coupling ($\gamma T \ll M_{a,b}\ll 1$) these density matrices at
time $t=t_0+\tau$ can be represented in the Bloch coordinates as
$z=A(t) \cos [\Omega t-\phi(t)]$, $y={\mbox Tr} (\sigma_y \rho)
=A(t)\sin [\Omega t-\phi(t)]$ with slowly changing amplitude $A(t)$
and phase $\phi(t)$:
    \begin{eqnarray}
&& \hspace{-0.8cm}    \dot{A}= -(A/T) [M_a \sin^2 (-\phi) +M_b
\sin^2 (\varphi-\phi)] -\gamma A/2,
       \label{A-dot} \\
&& \hspace{-0.8cm}  \dot \phi = (1/2T) [M_a \sin (-2\phi) + M_b\sin
(2\varphi-2\phi)] ,
    \label{phi-dot}\end{eqnarray}
where we assumed $\delta t_{a,b}\ll T$, so that periodic
instantaneous dephasings of magnitudes $M_a$ and $M_b$ happen at the
phases $-\phi$ and $\varphi-\phi$. Assume now $t_0=0$ [so that
$f_a(t_0)=1$] and choose $|i\rangle=|1\rangle$; then  $A(0)=1$ and
$\phi(0)=0$. For $|\varphi | \ll 1$ the solution of Eq.\
(\ref{phi-dot}) is simple, and $\phi$ saturates at
$\phi_{st}=\varphi M_b/(M_a+M_b)$ exponentially with time constant
$t_{\rm col}$. This value can be inserted into Eq.\ (\ref{A-dot})
because evolution of the amplitude $A$ due to measurement is much
slower than $t_{\rm col}^{-1}$, resulting in $A(t)=\exp[-(t/T)
\varphi^2 M_aM_b/(M_a+M_b)-\gamma t/2]$. It is easy to see that the
evolution starting with the state $|2\rangle$ leads to $\phi$
shifted by $\pi$, but the same $A(t)$, and the same contribution
into $K^{zz}(\tau,0)$ in Eq.\ (\ref{K-zz}). Calculating now
$S_{aa}(\omega)$ via $K^{zz}(\tau,0)$, and using approximation
$|\cos\phi_{st}|\approx 1$ because $|\phi_{st}|<|\varphi|\ll 1$, we
obtain the formula, coinciding with Eq.\ (\ref{S-nn}) with $\Gamma_S
= (1/2T)\varphi^2 M_aM_b/(M_a+M_b)+\gamma/4$. Calculation of
$S_{bb}(\omega )$ is fully similar, while to obtain $S_{ab}(\omega
)$ in the form (\ref{S-ab}) with the same switching rate $\Gamma_S$
we also need approximation $|\cos (\phi-\phi_{st})|\approx 1$.

   To account for small non-zero pulse widths $\delta
t_{a,b}$, we can still use Eq.\ (\ref{phi-dot}), but averaging over
$\phi$ within the pulse widths in Eq.\ (\ref{A-dot}) leads to the
extra factor $\exp [-(2\pi)^2(M_a\delta t_a^2+M_b\delta t_b^2)t/12
T^3]$ in $A(t)$ and corresponding increase of $\Gamma_S$. A small
frequency mismatch $\Delta\Omega =\Omega-\Omega_R$ would lead to the
extra term $\Delta \Omega$ in Eq.\ (\ref{phi-dot}), so that
$\phi_{st}$ becomes shifted by $\Delta \Omega \, T/(M_a +M_b)$. In
the case when the shift between the measurement combs is close to
the half-period, $\varphi$ should be obviously replaced by
$\varphi\pm \pi$. Taking into account these changes, we reproduce
Eqs.\ (\ref{S-nn}) and (\ref{S-ab}) with the switching rate
    \be
    \Gamma_S =   \frac{\tilde\varphi^2 M_aM_b +(\Delta \Omega \, T)^2}
{2T(M_a+M_b)} +\frac{M_a\delta t_a^2 +M_b\delta t_b^2}{6T^3/\pi^2}
+\frac{\gamma}{4}
    \label{Gamma-S}\ee
for $\tilde\varphi\ll 1$, where $\tilde\varphi =\min (|\varphi|,
|\varphi\pm\pi|)$. Assuming comparable measurement parameters for
both detectors, we see that the telegraph noise at zero frequency
greatly exceeds the shot noise contribution in Eq.\ (\ref{S-nn}) if
$\tilde\varphi \ll 1$, $\delta t_{a,b}\ll T$, $|\Delta \Omega| T \ll
M_{a,b}$, and $\gamma T \ll M_{a,b}$. This is the condition for the
validity of our analytical results.

\begin{figure}[tb]
  \centering
\includegraphics[width=7.0cm]{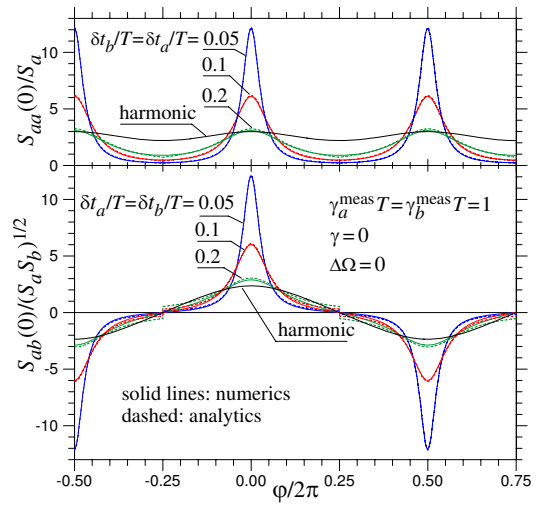}
  \caption{ (Color online) Numerical (solid lines) and analytical
(dashed lines) dependence of zero-frequency detector noise
$S_{aa}(0)$ and cross-noise $S_{ab}(0)$ on the phase shift $\varphi$
between the bias voltage combs for several values of the pulse width
$\delta t_{a,b}$, and also for the harmonic biasing. Almost complete
noise anticorrelation at $\varphi=\pm \pi$ indicate persistent Rabi
oscillations.}
  \label{fig2}
\end{figure}

    Figure 2 shows zero-frequency spectral densities $S_{aa}(0)$ and
$S_{ab}(0)$ as functions of the phase shift $\varphi$ for several
values of the pulse width $\delta t_{a,b}$, assuming negligible
$\Delta \Omega$ and $\gamma$. Solid lines are the numerical results
calculated via Eq.\ (\ref{K-zz}), while the dashed lines show
analytical results using Eqs.\ (\ref{S-nn}), (\ref{S-ab}), and
(\ref{Gamma-S}). Overall the analytics is very close to the
numerical results (almost coinciding), except for $S_{ab}(0)$ near
$\varphi =\pm \pi/2$, where the analytics is discontinuous because
of the sign change in (\ref{S-ab}). We normalize the noise
$S_{aa}(0)$ by the shot noise $S_a$ of constantly biased detector,
so that the shot noise contribution in this normalization is $\delta
t_a/T$. Similarly, the cross-noise $S_{ab}(0)$ is normalized by
$\sqrt{S_a S_b}$. The numerical results in Fig.\ 2 confirm almost
full correlation of the detector noises at $\varphi \approx 0$ and
almost full anticorrelation at $\varphi\approx\pm \pi$
\cite{antisym}. The peaks become higher and narrower for shorter
pulse durations $\delta t_{a,b}$. The results in the used
normalization are practically insensitive to the qubit-detector
coupling $\gamma^{\rm meas}_{a,b}$ (assuming $M_{a,b}\ll 1$).
Non-zero detuning $\Delta \Omega$ and/or extra dephasing $\gamma$
make the peaks in Fig.\ 2 lower, while not affecting their width;
this lowering is less significant for stronger coupling $\gamma^{\rm
ms}_{a,b}$. We have also checked numerically that the frequency
dependence of the noises at $\omega \ll \Omega$ is close to the
analytical results (\ref{S-nn}) and (\ref{S-ab}); extra peaks as
well as significant imaginary component of $S_{ab}(\omega)$ appear
at $\omega \approx \Omega$ and overtones of $\Omega$.

    Obviously, the analysis and results change only trivially
if $\Omega_R/\Omega$ is an integer or close to an integer.  In a
real experiment with QPC detectors the best measurement mode is to
apply two bias voltage pulses with opposite polarity per Rabi period
for each detector. In this case the average bias voltage is zero
that helps to keep zero bias between the pulses. The average current
in each detector is then also zero, simplifying the noise
measurement. For such mode $\delta t_{a,b}$ in Eqs.\ (\ref{S-nn})
and (\ref{S-ab}) should be replaced by $2\delta t_{a,b}$, while in
Eq.\ (\ref{Gamma-S}) the measurement strengths $M_{a,b}$ should be
doubled (no change for $\delta t_{a,b}$).

    Now let us discuss why experimental observation of the noise
dependence of Fig.\ 2 would indicate persistent Rabi oscillations.
Correlation of the noises for $\varphi \approx 0$ could be
alternatively explained by the qubit localization in either state
$|1\rangle$ or $|2\rangle$. However, the anticorrelation for
$\varphi \approx \pm\pi$ is possible only if the qubit oscillates
persistently. Moreover, these oscillations should be synchronized
with the measurement combs, because for persistent Rabi oscillations
with a random phase one would expect dependence $S_{ab}(0)\propto
\cos \varphi$, which is very different from the peaked dependence in
Fig.\ 2.  One may also worry that the noise dependence of Fig.\ 2
could be alternatively explained by the driven Rabi oscillations
(between the energy eigenstates) caused by presence of a voltage
with resonant frequency. However, both energy eigenstates produce no
signal in the detectors; therefore the driven Rabi oscillations
could only reduce the discussed noise correlation and cannot be used
for an alternative explanation (notice also that both stable phases
of the persistent oscillations are insensitive to the microwave
drive $\propto \cos \Omega_R t$). Unfortunately, measurement of only
zero-frequency noise is insufficient to demonstrate $\sim$100\%
amplitude of the persistent Rabi oscillations (observed in
\cite{Bertet}). However, if the switching rate $\Gamma_S$ can be
measured either in frequency or time domain, then by comparing
$S_{ab}(0)$ with Eq.\ (\ref{S-ab}) one can check the amplitude of
synchronized oscillations.

    If the stroboscopic biasing is replaced by harmonic biasing
\cite{Weidenmuller}: $f_a= \cos(\Omega t)$, $f_b=\cos (\Omega
t-\varphi)$, then $S_{ab}(0)$ still depends on the phase shift
$\varphi$ (see Fig.\ 2); however, there are no more peaks and the
noise magnitude is relatively small. The numerical results at weak
coupling can be fitted as $S_{ab}(0)=1.18 \Delta I_a \Delta I_b \cos
\varphi/(\gamma_{a}^{\rm meas}+\gamma_b^{\rm meas})$ (actually, the
$\varphi$-dependence is slightly more peaked than $\cos\varphi$).

    In our analysis we have neglected the noise from amplifiers, which
can be simply added and is not expected to depend on $\varphi$. The
main effect of the neglected thermal noise in the QPCs is a small
contribution to the dephasing $\gamma$ between the pulses. A weak
energy relaxation in the qubit can also be easily taken into
account.

    For numerical estimates let us assume QPCs with $I_{a,b}\simeq 100$ nA,
$\Delta I_{a,b}/I_{a,b}\simeq 0.1$, symmetric biasing with $\delta
t_{a,b}/T\simeq 0.1$, and Rabi frequency $\Omega_R/2\pi\simeq 2$
GHz. Then the collapse (``attraction'') time $t_{\rm col}\simeq 2$
ns is few Rabi periods, while the switching rate is $\Gamma_S\simeq
\tilde\varphi^2 /15\, \mbox{ns}+ 1/120 \, \mbox{ns}+ (\Delta
\Omega/\Omega)^2/6 \, \mbox{ps}+ \gamma/4$. Therefore we need the
dephasing time $T_2=1/\gamma$ to be longer than only few ns to have
significant correlated telegraph noise, and its ratio to the shot
noise contribution for $\tilde\varphi=\Delta\Omega=0$ is crudely
$\min (60, T_2/0.5 \mbox{ns})$ (5 times smaller for the
normalization of Fig.\ 2). These figures show that the experiment is
doable using the present-day semiconductor technology
\cite{Rabi-semicond,semicond}. The experiment can also be realized
with the superconducting qubit setup of Ref.\ \cite{Bertet}. In
comparison with the experiment \cite{Bertet} it would also
demonstrate partial synchronization of the persistent Rabi
oscillations without use of a much more complicated quantum
feedback.

The author thanks Rusko Ruskov for useful discussions. This work was
supported by NSA and IARPA under ARO grant W911NF-08-1-0336.




\begin{thebibliography}{99}
\bibitem{antisym} Actually, geometrically antisymmetric qubit
coupling to the QPC detectors in Fig.\ 1 flips the sign of the noise
correlation. We neglect this effect (because it is absent in a
superconducting setup) by assuming positive $\Delta I_a$ and $\Delta
I_b$ in Eq.\ (\ref{I-n}) by defining proper current directions.
\end{thebibliography}
\end{document}